\documentclass[%
 reprint,
 amsmath,amssymb,
 aps,
]{revtex4-2}

\usepackage{float}
\usepackage{mathtools}
\usepackage{amsmath}
\usepackage{amsthm}
\usepackage{graphicx}
\usepackage{dcolumn}
\usepackage{bm}

\begin{document}


\title{JT Gravity from Partial Reduction and Defect Extremal Surface}
\author{Feiyu Deng}
\email{fydeng20@fudan.edu.cn}
\author{Yu-Sen An}
\email{anyusen@fudan.edu.cn}

\affiliation{Department of Physics and Center for Field Theory and Particle Physics, Fudan University, Shanghai 200433, China}

\author{Yang Zhou}
\email{yang\_zhou@fudan.edu.cn}
\affiliation{Department of Physics and Center for Field Theory and Particle Physics, Fudan University, Shanghai 200433, China}
\affiliation{Peng Huanwu Center for Fundamental Theory, Hefei, Anhui 230026, China}
\date{\today}

\begin{abstract}
We propose the three-dimensional bulk dual for Jackiw-Teitelboim gravity coupled with CFT$_2$ bath based on partial reduction. The bulk dual is classical AdS gravity with a defect brane which has small fluctuation in transverse direction. We derive full Jackiw-Teitelboim gravity action by considering the transverse fluctuation as a dilaton field. We demonstrate that the fine grained entropy computed from island formula precisely agrees with that computed from defect extremal surface. Our construction provides a Lorentzian higher dimensional dual for Jackiw-Teitelboim gravity and therefore offers a framework to study problems such as black hole information paradox as well as gravity/ensemble duality.

\end{abstract}

\maketitle

\subsection{Introduction}
Significant progress has been made in recent understanding of black hole information paradox~\cite{Hawking:1976ra}. In particular the island formula~\cite{Penington:2019npb,Almheiri:2019psf,Almheiri:2019hni} for the von Neumann entropy of Hawking radiation~\cite{Hawking:1975vcx} gives Page curve~\cite{Page:1993wv} and therefore maintains unitarity. The development relies on the quantum extremal surface formula~\cite{Engelhardt:2014gca} for the fine grained entropy, which was based on the quantum corrected Ryu-Takayanagi formula in computing holographic entanglement entropy~\cite{Ryu:2006bv,Hubeny:2007xt,Faulkner:2013ana}. To justify the island formula of von Neumann entropy,  one can employ replica trick and perform explicit gravitational path integral computation in lower dimensional systems. In particular the Jackiw-Teitelboim (JT) gravity coupled to a quantum bath provides a solvable model to explore the information transfer for black hole in two dimensions~\cite{Almheiri:2019qdq,Penington:2019kki}. It has been found that the island contribution corresponds to Euclidean replica wormholes \cite{Penington:2019kki}, which may cause the factorization puzzle~\cite{Witten:1999xp,Maldacena:2004rf}. 
It is still quite interesting to explore the Lorentzian counterparts for the replica wormholes. There are many applications and generalizations of island formula, including the application in cosmology \cite{Chen:2020tes,Hartman:2020khs} and the generalization to asymptotically flat space \cite{Hartman:2020swn,Krishnan:2020oun}.
For other related works, see \cite{Chen:2019iro,Chen:2020wiq,Sully:2020pza,Hollowood:2020cou,Geng:2020qvw,Chen:2019uhq,Li:2020ceg,Chandrasekaran:2020qtn,Dong:2020uxp,Chen:2020jvn,Balasubramanian:2020xqf,Chen:2020uac,Chen:2020hmv,Ling:2020laa,Harlow:2020bee,Hernandez:2020nem,Chen:2020ojn,Akal:2020wfl,Rozali:2019day,Chen:2020hmv,Suzuki:2022xwv,Bousso:2022gth,Miyaji:2021lcq,Neuenfeld:2021bsb,Verheijden:2021yrb,Akal:2020twv}. On the other hand, AdS gravity with a defect brane provides a natural holographic framework to study lower dimensional gravity coupled to a bath. There the holographic dual of von Neumann entropy is Ryu-Takayangai surface or Hubeny-Rangamani-Takayanagi surface and therefore Lorentzian. The defect brane model can also be generalized to higher dimensions \cite{Almheiri:2019psy}.

In \cite{Deng:2020ent}, it has been shown that the holographic counterpart of island formula is Defect Extremal Surface (DES) formula, which is the defect corrected Ryu-Takayanagi formula. The defect brane model is based on $\text{AdS}_{\text{d+1}}/\text{BCFT}_{\text{d}}$ \cite{Takayanagi:2011zk}, where the defect brane with a constant tension in the AdS bulk has $\text{AdS}_{\text{d}}$ geometry and there are also quantum degrees of freedom localized on it. By doing partial dimension reduction between zero tension brane and constant tension brane, one can obtain the brane world gravity, which is coupled to a CFT bath through boundary conditions.
In three dimensional bulk, the simple dimension reduction leads to a two dimensional topological gravity on the brane ~\cite{Deng:2020ent}. It has been further demonstrated that entanglement entropy and reflected entropy computed from DES formula agrees with island formula precisely~\cite{Li:2021dmf}.
See \cite{Chu:2021gdb,Wang:2021xih,Shao:2022gpg,Basu:2022reu} for further works along this line.
To fully reconcile the path integral approach and holographic approach, one may ask whether we can obtain JT gravity, which has a dilaton field, directly from dimension reduction in the defect brane model. If yes, what would be the holographic counterpart of the island formula for JT gravity coupled with CFT bath?


In this letter we answer these questions.
We show that for the 3d AdS bulk with a defect brane, by considering the small transverse fluctuation, one can derive the full JT gravity action from partial reduction between zero tension brane and finite tension brane. In particular, the transverse fluctuation becomes the dilaton field on the brane world. For the remaining part of the bulk one can use standard AdS/CFT and obtain CFT$_2$ bath on the asymptotic boundary. Eventually we obtain a 2d JT gravity coupled with CFT$_2$ bath. The fact that the transverse fluctuation is small allows us to ignore higher order contributions in 2d action and obtain precisely the full 2d JT action including boundary term.
We therefore obtain a higher dimensional dual for JT gravity coupled to a bath.

To support the 3d/2d duality, we compute the fined-grained entropy both from defect extremal surface formula in the bulk and from the boundary island formula. We find that the extremal conditions are consistent and the fined-grained entropy agree with each other precisely for small transverse fluctuation. We consider this agreement as strong evidence to support that JT gravity coupled to CFT bath can be dual to semi-classical 3d gravity with a fluctuating brane.

\subsection{Review of the model}
We consider $\rm{AdS_3}/\rm{BCFT_2}$ with the action given by 
\begin{equation}\begin{split}
I&=\frac{1}{16 \pi G_{ N}} \int_{ N} \sqrt{-g}(R-2 \Lambda)\\
&+\frac{1}{8 \pi G_{ N}} \int_{ M}  \sqrt{-\gamma} K^{(\gamma)}\\
&+\frac{1}{8 \pi G_{ N}} \int_{ Q} \sqrt{-h}K^{(h)}\\
&+I_{ Q}+I_{ P}\ ,
\end{split}\end{equation}
where $N$ denotes the bulk AdS spacetime, $M$ denotes the asymptotic boundary where the Dirichlet boundary condition is imposed and $Q$ the brane where Neumann boundary condition is imposed. $ I_{ Q}$ is the action for matter fields constrained on $Q$ and $I_{P}$ is the counter term on the tip $P$. By varying this action, the Neumann boundary condition on $Q$ becomes
\begin{equation}
K_{a b}^{(h)}-h_{a b} K^{(h)}=8\pi G_{ N}T_{ab}\ \label{NBC} ,
\end{equation}
where $T_{ab}=-\frac{2}{\sqrt{-h}} \frac{\delta I_{Q}}{\delta h^{a b}}$ is the stress energy tensor coming from the variation of matter action.

Here we consider the bulk to be 3 dimensional and the brane $Q$ is thus two dimensional. There are two sets of useful coordinates: $(t,x,z)$ and $(t,\rho,y)$. Their relation is
\begin{equation}
z=-y / \cosh \frac{\rho}{l}\ , \quad x=y \tanh \frac{\rho}{l}\,
\end{equation}
and the bulk metric can be written in terms of either one
\begin{equation}\begin{split}
d s^{2}_N
&=\frac{l^2}{z^2}(-dt^2+dz^2+dx^2)\\
&=d \rho^{2}+l^2\cosh ^{2} \frac{\rho}{l} \cdot \frac{-d t^{2}+d y^{2}}{y^{2}},
\end{split}\end{equation}
where $l$ is the AdS radius. It is also useful to introduce polar coordinate $\theta$ with $\frac{1}{\cos (\theta)}=\cosh \left(\frac{\rho}{l}\right)$.

\begin{figure}[H]
	\centering
	\includegraphics[width=7.7cm,height=4.7cm]{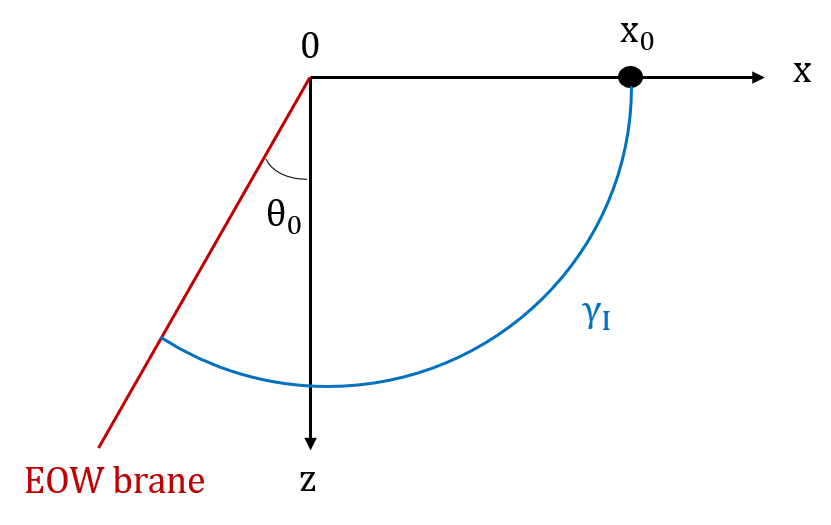}\\
	\caption{\label{AB} The set up of AdS/BCFT where the brane tension is a constant.}
\end{figure}

Consider the matter action on the brane $Q$ of the form
\begin{equation}
I_{Q}=-\frac{1}{8 \pi G_{N}} \int_{Q} \sqrt{-h}T,
\end{equation}
where $T$ is a constant tension. By solving (\ref{NBC}), one can determine that the brane is located at $\rho=\rho_0=\text{arctanh}\sin\theta_0$, where $\rho_0$ is a positive constant. See FIG. \ref{AB} for an illustration.
 The tension $T$ is found to be 
\begin{equation}\label{T}
	T=\frac{\tanh\frac{\rho_0}{l}}{l}.
\end{equation}

For an interval $I:=[0,x_0]$ in BCFT, the entanglement entropy can be computed holographically using RT formula. As shown in FIG.\ref{AB}, the minimal surface denoted by $\gamma_I$ terminates on a point on the brane which can be determined by extremization. The entanglement entropy is
\begin{equation}\begin{split}
	S_{I}&=\frac{\operatorname{Area}\left(\gamma_{I}\right)}{4 G_{N}}=\frac{c}{6} \log \frac{2x_0}{\epsilon}+\frac{c}{6}\rho_0\\
	&=\frac{c}{6} \log \frac{2x_0}{\epsilon}+\frac{c}{6} \operatorname{arctanh}(\sin \theta_0),
\end{split}\end{equation}
where $c$ is the CFT central charge and $\epsilon$ is the UV cut off.

In \cite{Deng:2020ent}, the authors improved AdS$_3$/CFT$_2$ by adding CFT matter localized on the brane. Notice that ${\rm{AdS}}_2$ is a maximally symmetric space, the vacuum one point function of the CFT stress tensor takes the form
\begin{equation}
\label{st}
\langle T_{ab}\rangle_{\text{AdS}_2}=\chi h_{ab}\ ,
\end{equation}
which contributes to Neumann boundary condition (\ref{NBC}). Because of the entangled quantum matter on the defect brane, we should add the defect contribution to the ordinary holographic entanglement entropy. The improved holographic formula is therefore called Defect Extremal Surface (DES) formula \cite{Deng:2020ent}.
\subsection{JT gravity from dimensional reduction}

\begin{figure}
	\centering
	\includegraphics[width=7cm,height=9cm]{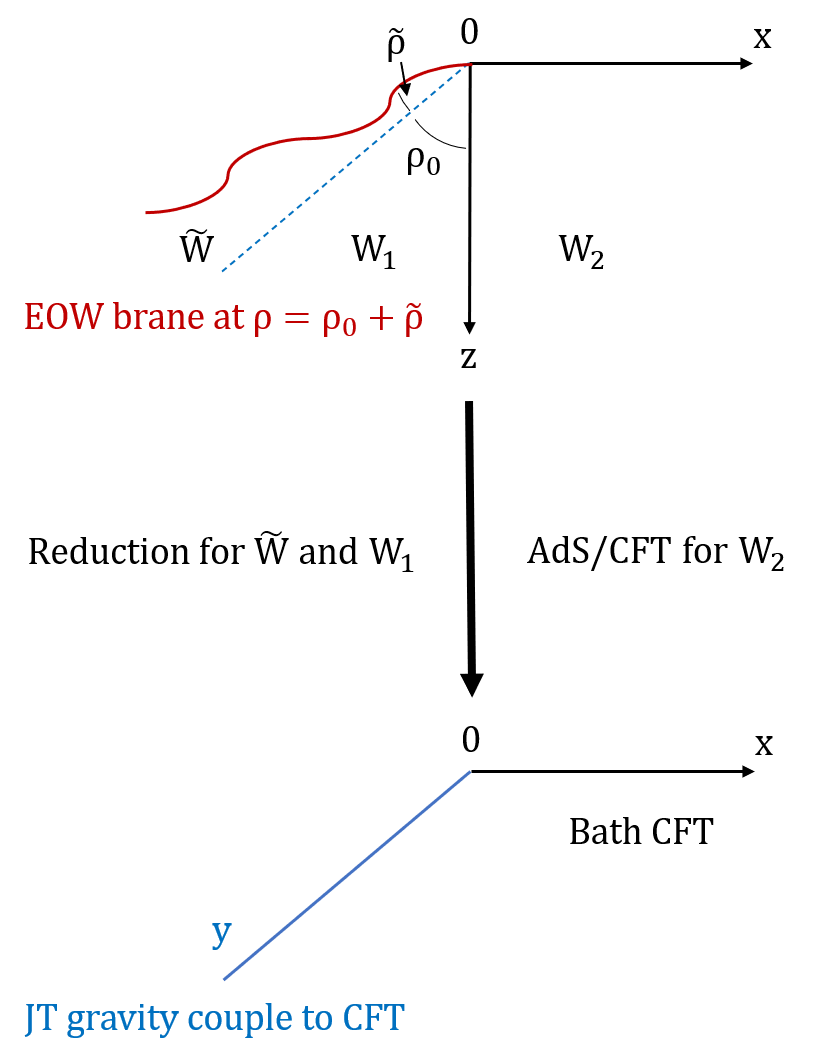}\\
	\caption{\label{parred} Effective description from Partial Dimension Reduction plus AdS/CFT.}
\end{figure}

Now we derive JT gravity from dimension reduction of 3d AdS gravity action.
Under the metric ansatz
\begin{equation}
    ds^{2}=g_{\mu\nu}dx^{\mu}dx^{\nu}=d\rho^{2}+l^{2}\cosh^{2}\frac{\rho}{l} {\tilde{h}}_{ab}dx^{a}dx^{b},
\end{equation}
one can perform partial dimensional reduction for wedge $W_1+\tilde{W}$ by integrating out the $\rho$ direction as shown in FIG. \ref{parred}. The bulk action becomes
	\begin{equation}\begin{split}\label{EH2}
		&\frac{1}{16 \pi G_{N}} \int_{W_1+\tilde{W}} \sqrt{-g}(R-2 \Lambda) \\
		&=\int\frac{\rho_0+\tilde{\rho}}{16 \pi G_{N}} \sqrt{-g^{(2)}} R^{(2)}\\
		&-\int\frac{1}{16 \pi G_{N}}\frac{\sinh (\frac{2\rho}{l})}{l\cosh^2\frac{\rho_{0}}{l}} \sqrt{-g^{(2)}},
		\end{split}
	\end{equation}
where $\Lambda=-\frac{2}{l^2}$ and
\begin{equation}
    g_{ab}^{(2)}=l^{2}{\rm{cosh}}^{2}\frac{\rho_{0}}{l} {\tilde{h}}_{ab}.
\end{equation}
The precise reduction of bulk Ricci scalar
\begin{equation}
	\sqrt{-g}R=\sqrt{-g^{(2)}}\left [R^{(2)}-\frac{2(3\cosh^2\frac{\rho}{l}-1)}{l^2\cosh^2\frac{\rho_0}{l}}\right]
\end{equation}
has been used in eq.(\ref{EH2}).
Notice that the brane $Q$ is located at 
\begin{equation}
\rho=\rho_0+\tilde{\rho},
\end{equation}
where $\tilde{\rho}$ is a small fluctuation away from $\rho_0$, i.e.$\frac{\tilde{\rho}}{\rho_{0}}\ll 1 $. The fluctuation $\tilde{\rho}$ is a function of the brane world coordinates and therefore should be treated as a field on the brane.

Next we consider Gibbons-Hawking term and the brane tension term. The extrinsic curvature of the brane is 
\begin{equation}
	K_{ab}=\frac{\tanh(\frac{\rho_0+\tilde{\rho}}{l})}{l}{h}_{a b},
\end{equation}
where ${h}_{ab}$ is the induced metric on the brane
\begin{equation}
    {h}_{ab}=l^{2}{\rm{cosh}}^{2}\frac{\rho}{l}\tilde{h}_{ab}.
\end{equation}
The tension of the brane remains a constant which equals to the tension (\ref{T}) since the fluctuation of the $\rho$ coordinate will not affect the intrinsic brane tension.
Then Gibbons-Hawking term plus the brane tension term
is given by
\begin{equation}\begin{split}\label{GH}
	&\frac{1}{8 \pi G_{N}}\int_Q \sqrt{-{h}}(K-T)\\
	&=\frac{1}{8 \pi G_{N}}\int \frac{\sinh(\frac{2\rho_0+2\tilde{\rho}}{l})}{l\cosh^2\frac{\rho_0}{l}}\sqrt{-g^{(2)}}\\
	&-\frac{1}{8 \pi G_{N}}\int\frac{\tanh\frac{\rho_0}{l}\cosh^2(\frac{\rho_0+\tilde{\rho}}{l})}{l\cosh^2\frac{\rho_0}{l}}\sqrt{-g^{(2)}}.
\end{split}\end{equation}
By adding (\ref{EH2}) and (\ref{GH}) together and expanding with small $\tilde{\rho}/\rho_0$, we get the action of the 2d effective theory after partial dimension reduction
\begin{equation}\begin{split}
	I_{tot}&=
	\frac{\rho_0}{16 \pi G_{N}}\int \sqrt{-g^{(2)}}\frac{\tilde{\rho}}{\rho_0}\left(R^{(2)}+\frac{2}{l^2\cosh^2\frac{\rho_0}{l}}\right)\\
	&+\frac{\rho_0}{16 \pi G_{N}}\int \sqrt{-g^{(2)}} R^{(2)}+\mathcal{O}(\frac{\tilde{\rho}^2}{{\rho_0}^2}).
\end{split}\end{equation}
Neglecting $\mathcal{O}(\frac{\tilde{\rho}^2}{{\rho_0}^2})$ terms, we see that the action is precisely the JT action, and $\frac{\tilde{\rho}}{\rho_0}$ is identified as the dilaton field in JT gravity. If we vary with respect to $\frac{\tilde{\rho}}{\rho_0}$, we get the scalar curvature 
\begin{equation}
	R^{(2)}=-\frac{2}{l^2\cosh^2\frac{\rho_0}{l}},
\end{equation}
which gives the correct scalar curvature of the ${\rm{AdS}}_2$ brane world.

\begin{figure}
	\centering
	\includegraphics[width=5cm,height=4cm]{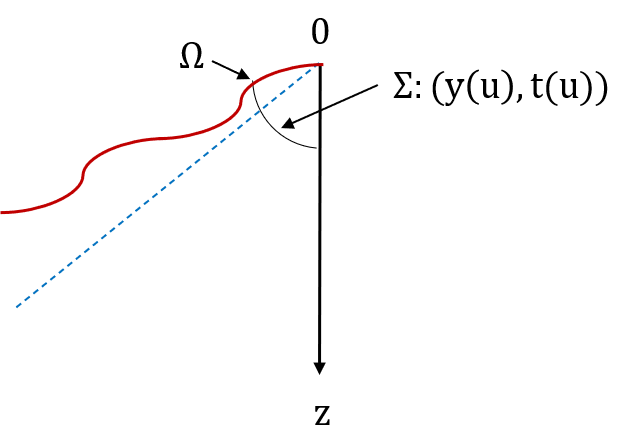}\\
	\caption{\label{UV} Dynamical UV cut off for the wedge $W_1 + \tilde{W}$.}
\end{figure}

To fully recover the JT action, we also need to reproduce the boundary term. Now we show that the JT boundary action can be obtained by doing dimension reduction from the Gibbons-Hawking term on the bulk cutoff surface. 
\footnote{This part of calculation is closely related to the calculation in \cite{Akal:2020wfl}.}. As shown in FIG. \ref{UV}, the cutoff surface is denoted as $\Sigma$.
Near the asymptotic boundary, the metric is taken to be 
\begin{equation}
    ds^{2}=d\rho^{2}+l^{2}\cosh^{2}\frac{\rho}{l}\frac{-dt^{2}+dy^{2}}{y^{2}}.
\end{equation}
Consider a generic cutoff surface $\Sigma$ parameterized by
\begin{equation}
	(y,t)=(y(u),t(u)),
\end{equation}
where $u$ is the time in the cutoff boundary.  
The induced metric at cutoff surface is
\begin{equation} \label{uv1}
	ds^{2}=d\rho^{2}+l^{2}\cosh^{2}(\frac{\rho}{l})(\frac{y'^{2}du^{2}-t'^{2}du^{2}}{y^{2}})
\end{equation}
where we denote $y'=\frac{dy}{du}$, $t'=\frac{dt}{du}$. One can compute the extrinsic curvature of $\Sigma$ by $K=g^{\mu\nu}\nabla_{\mu}n_{\nu}$ and the result is 
\begin{equation}
	K_{\Sigma}=\frac{1}{l \cosh\frac{\rho}{l}} \frac{t'^{3}+y y't''-t'y'^{2}-t'y y''}{(t'^{2}-y'^{2})^{3/2}}=\frac{1}{l \cosh\frac{\rho}{l}} K_{\Omega}
\end{equation}
where $K_{\Omega}=\frac{t'^{3}+y y't''-t'y'^{2}-t'y y''}{(t'^{2}-y'^{2})^{3/2}}$ is the extrinsic curvature of the boundary $\Omega$ which is the intersection between $\Sigma$ and EOW brane. 
By doing dimension reduction of Gibbons-Hawking term on $\Sigma$, the action on $\Omega$ is obtained
\begin{equation}
\begin{split}
	I_{bJT}&=\frac{1}{8 \pi G_{N}} \int_{\Sigma} \sqrt{-\gamma} K_{\Sigma}\\
	&=\frac{\rho_{0}}{8\pi G_{N}}\int du \sqrt{\frac{t'^{2}-y'^{2}}{y^{2}}}(1+\frac{\tilde{\rho}}{\rho_{0}}|_{\mathrm{bdy}})  K_{\Omega} ,
\end{split}
\end{equation}
This is precisely the boundary term of JT gravity when $\frac{\tilde{\rho}}{\rho_{0}}$ is identified as the dilaton field.
At the intersection $\Omega$, employing the same trick in \cite{Maldacena:2016upp} one can fix the induced metric
\begin{equation} \label{uv2}
    g|_{\mathrm{bdy}}=-\frac{l^{2}\cosh^{2}(\frac{\rho}{l})}{\epsilon^{2}}.
\end{equation}
By identifying the induced metric in (\ref{uv1}) to (\ref{uv2}), $y$ can be solved to the leading order in $\epsilon$, $y=\epsilon t'+O(\epsilon^{2})$.
By computing $K_{\Omega}$ directly, we find 
\begin{equation}
    K_{\Omega}=1-\epsilon^{2}{\rm{Sch}}(t(u),u).
\end{equation}
If we neglect the field independent divergent term, the boundary term is a Schwarzian \footnote{In \cite{Akal:2020wfl}, the authors also obtained Schwarzian theory by doing dimension reduction for the UV cutoff given by $y=g(t)$.}.
We leave the computation details to appendix \ref{shw}.

Thus after doing the partial dimension reduction for wedge $W_1+\tilde{W}$ and using standard AdS/CFT for wedge $W_2$, we obtain the 2d effective theory to be a JT gravity together with the brane CFT, glued to a non-gravitational CFT bath through transparent boundary conditions, which is precisely the original set up to motivate the island formula. See FIG. \ref{parred} for an illustration of this procedure.

\subsection{Entanglement entropy for an interval}
\subsubsection{Entanglement entropy from island formula}
Since we have obtained the 2d effective description in terms of a JT gravity glued with a CFT bath, we can use island formula to calculate the von Neumann entropy for an interval $[0,L]$ on the asymptotic boundary
\begin{equation}
S=\min _{X}\left\{\operatorname{ext}_{X}\left[\frac{\operatorname{Area}(X)}{4 G_{N}}+S_{\text {semi-cl }}\left(\Sigma_{X}\right)\right]\right\},
\end{equation}
where $X=\partial I$ and $I$ is the island, $\Sigma_X$ is the associated region including the island and $S_{\rm{semi-cl}}(\Sigma_X)$ is the von-Neumann entropy of the quantum fields on $\Sigma_X$ in the semi-classical description.
Introducing the 2d Newton constant
\begin{equation}
	G_N^{(2)}=\frac{G_N}{\rho_0},
\end{equation}
the action of 2d effective theory becomes
\begin{equation}\begin{split}
	I_{\text{2d}}&=\frac{1}{16 \pi G_{N}^{(2)}}\int \sqrt{-g^{(2)}} R^{(2)}\\
	&+\frac{1}{16 \pi G_{N}^{(2)}}\int \sqrt{-g^{(2)}}\frac{\tilde{\rho}}{\rho_0}\left(R^{(2)}+\frac{2}{l^2\cosh^2\frac{\rho_0}{l}}\right)\\
	&+I_{\text{CFT}}.
\end{split}
\end{equation}
Varying with respect to the 2d metric one can solve the dilaton to be 
\begin{equation}\label{JT}
	\frac{\tilde{\rho}}{\rho_0}=-\frac{\bar{\phi}_r}{y}.
\end{equation}
Now we employ island formula to calculate the entropy of a single interval $[0,L]$. For simplicity we focus on static cases. The generalized entropy is given by 
\begin{equation}
	\begin{aligned}
		S_{\text{gen}}(a)&=S_{\text {area }}(y=-a)+S_{\text {matter }}([-a, L]) \\
		&=\frac{1}{4G_N^{(2)}}(1+\frac{\tilde{\rho}}{\rho_0})+\frac{c}{6} \log \frac{(L+a)^{2} l}{a \cos \theta_{0} \epsilon \epsilon_{y}}.\\
	\end{aligned}
\end{equation}
The extremization condition $\partial_a	S_{\text{gen}}(a)=0$ determines the boundary of island $a$ to be 
\begin{equation}\label{aIS}
a=\frac{L}{2}\left(\sqrt{36\mu^{2}+36\mu+1}+6\mu+1\right),
\end{equation}
where $\mu$ is defined as $\mu=\frac{\rho_0\bar{\phi}_r}{6lL}$, which can also be expressed as $\frac{1}{6}\cdot\frac{\tilde{\rho}(a)}{l}\cdot\frac{a}{L}$ because of eq.(\ref{JT}). It characterizes the amplitude of the brane transverse fluctuation at location $a$, since $\frac{a}{L}$ measures the distance from the origin and $\frac{\tilde{\rho}}{l}$ measures the fluctuation in angular direction. 
In fact one can work out a simple relation between extremal point $a$ and $\tilde{\rho}(a)$
\begin{equation}
\frac{a}{L}=\frac{1+\frac{\tilde{\rho}(a)}{l}}{1-\frac{\tilde{\rho}(a)}{l}}.
\end{equation}

By plugging (\ref{aIS}) into $S_{\text{gen}}(a)$ one can get the entropy $S_{\text{island}}$ computed from island formula. 

\subsubsection{Entanglement entropy from bulk DES}
The defect extremal surface formula is the defect corrected Ryu-Takayanagi formula. For defect $D$ in the AdS bulk, the entanglement entropy is computed following
\begin{equation}\begin{split}
	S_{\mathrm{DES}}&=\min _{\Gamma}\left\{\operatorname{ext}_{\Gamma,X}\left[\frac{\operatorname{Area}(\Gamma)}{4 G_{N}}+S_{\mathrm{defect}}[D]\right]\right\} ,
\end{split}\end{equation}
where $X=\Gamma \cap D$ and $\Gamma$ is the codimension two RT surface. We consider the interval $[0,L]$ on the asymptotic boundary and use DES formula to calculate its entropy.
First we compute the RT surface $\frac{\operatorname{Area}(\Gamma)}{4 G_{N}}$ by using embedding coordinates. Considering $\frac{\tilde{\rho}}{\rho_{0}}\ll 1 $, the leading order result is
\begin{equation}\begin{split}
    \frac{\text{Area}(\Gamma)}{4G_{N}}&=\frac{l}{4G_{N}}\operatorname{arccosh}\left[\frac{(L+a\sin\theta_0)^2+a^2\cos^2\theta_0}{2a \cos\theta_0 \epsilon}\right]\\
    &+\frac{\rho_{0}\bar{\phi}_{r}}{4G_{N}a}.
\end{split}\end{equation}
We leave the details to appendix \ref{RT}.
The brane CFT also contributes to the entanglement entropy
\begin{equation}
    S_{\text{defect}}(D)=\frac{c^{\prime}}{6} \log \frac{2 l}{\epsilon_{y} \cos \theta_{0}}.
\end{equation}
Then the total generalized entropy is
\begin{equation}\begin{split}
	S_{\text{gen}}(a)
	&=\frac{c}{6}\operatorname{log}\left[\frac{(L+a\sin\theta_0)^2+a^2\cos^2\theta_0}{a
		\cos\theta_0\epsilon}\right]\\
	&+\frac{\rho_0\bar{\phi}_r}{4G_N a}+\frac{c^{\prime}}{6} \log \frac{2 l}{\epsilon_{y} \cos \theta_{0}},
\end{split}\end{equation}
where $c=\frac{3l}{2G_N}$ is used and $c'$ is the central charge of the brane CFT. To compare with the result obtained from boundary island formula, we take $c'=c$.

From $\partial_a S_{gen}(a)=0$, one obtains the extremal position of $a$ to be
\begin{equation}\begin{aligned}\label{aDES}
a&=\frac{ (3)^{\frac{2}{3}}L \left(12 \mu^2+12 \mu \sin\theta_0 +1\right)}{3\nu}\\
&+2 \mu L+\frac{L \nu}{3^{\frac{2}{3}}},
\end{aligned}\end{equation}
where
\begin{equation}\begin{split}
\nu&=\sqrt[3]{72  \mu^3+\frac{1}{6} \sqrt{\gamma}+108  \mu^2  \sin\theta_0 +36  \mu }\\
\gamma&=46656 \mu^2 \left(2  \mu^2+3  \mu    \sin\theta_0 +1\right)^2\\
&-108 \left(12  \mu^2+12  \mu   \sin\theta_0 +1\right)^3.
\end{split}\end{equation}
Plugging (\ref{aDES}) into $S_{\text{gen}}(a)$, we have the entropy result calculated from DES.
\subsubsection{Comparison between DES and island formula}\label{Comparison}
Now we compare the entropy result computed from DES and that from island formula. 

We first consider $\rho_0/l\gg 1$ with the small fluctuation condition $\tilde{\rho}/\rho_0\ll 1$ satisfied. This is the limit that the brane is nearly parallel to the asymptotic boundary. In this limit, the extremal point (\ref{aIS}) and (\ref{aDES}) coincide with each other and both DES and island formula give the same entropy. To see this, let us fix $\mu$ and set $\theta_0=\frac{\pi}{2}-\omega$ \footnote{Although $\rho_0$ is very large, $\mu$ can be fixed.}. Expanding the extremal point and the entropy around $\omega=0$, we get
\begin{equation}\begin{split}
	a_{\text{DES}}&=\frac{L}{2}\left(\sqrt{36\mu^{2}+36\mu+1}+6\mu+1\right)+\mathcal{O}(\omega^2)\\
	&=a_{\text{island}}+\mathcal{O}(\omega^2),
\end{split}\end{equation}
and 
\begin{equation}
	\begin{split}
			S_{\text{DES}}
		&=\frac{2c\mu}{6 \mu+1 +\eta}+\frac{c}{6}\log \frac{ (6\mu+3 +\eta)^2L}{(6\mu+1+\eta)\epsilon}\\
		&+\frac{c}{6}\log\frac{l}{\omega^{2}\epsilon_y}+\mathcal{O}(\omega^2)\\
		&=S_{\text{island}}+\mathcal{O}(\omega^2),
	\end{split}
\end{equation}
 where 
 \begin{equation}
 	\eta=\sqrt{36 \mu^2+36\mu+1}.
 \end{equation}

Furthermore, we can consider a generic brane location, i.e. $\rho_0/l$ is finite. By the small fluctuation condition $\frac{\tilde{\rho}}{\rho_0}\ll 1$, we have that $\tilde{\rho}/l$ is small, which implies that $\mu=\frac{1}{6}\cdot\frac{\tilde{\rho}}{l}\cdot\frac{a}{L}\ll1$ provided that $a/L$ is order one. In this limit we can expand with small $\mu$. The extremal point of bulk DES becomes
\begin{equation}
	a=L+6(1+\sin\theta_0)L\mu+\mathcal{O}(\mu^2),
\end{equation}
and the extremal point of island formula is
\begin{equation}
	a=L+12L\mu+\mathcal{O}(\mu^2).
\end{equation}
The entropy from bulk DES is
\begin{equation}
	\begin{split}
		S_{\text{DES}}&=\frac{c}{6} \log \frac{2L}{\epsilon}+\frac{c}{6} \log\frac{1+\sin\theta_0}{\cos\theta_0}\\
		&+\frac{c}{6} \log \frac{2l}{\epsilon_y \cos\theta_0}+c\mu+\mathcal{O}(\mu^2),
	\end{split}
\end{equation}
while the entropy from boundary island formula is
\begin{equation}
	\begin{split}
		S_{\text{island}}&=\frac{c}{6} \log \frac{2L}{\epsilon}
		+\frac{c}{6} \operatorname{arctanh}(\sin \theta_0)\\
		&+\frac{c}{6} \log \frac{2l}{\epsilon_y \cos\theta_0}+c\mu+\mathcal{O}(\mu^2).
	\end{split}
\end{equation}
Comparing the above two results, we find that the entanglement entropy for single interval $[0,L]$ from DES and from island formula precisely match for small brane fluctuations.

\subsection{Conclusion and Discussion}
In this paper we constructed the three-dimensional bulk dual for Jackiw-Teitelboim gravity coupled to CFT$_2$ bath based on partial reduction. The bulk dual is classical AdS gravity with a defect brane which has small fluctuation in transverse direction. We obtain full Jackiw-Teitelboim gravity action by identifying the transverse fluctuation as a dilaton field on brane world. We further demonstrated that the fine grained entropy computed from island formula precisely agrees with that computed from defect extremal surface. There are a few interesting future questions: First, using our construction to understand JT gravity/ensemble relation. In our construction the JT gravity from partial reduction is dual to a defect theory in asymptotic boundary and it would be interesting to check the relation between this defect theory and the ensemble of quantum mechanics discussed in \cite{Saad:2019lba}. Second, generalize our construction to higher dimensions. In higher dimensions, the dilaton field is known as radion field in the original Randall-Sundrum model \cite{Randall:1999ee}. It is quite interesting to work out the full brane world theory including the dilaton field from partial reduction. Last but not least, it is interesting to test our construction by other entanglement measures and to explore the physical implications of the dilaton in brane world cosmology~\cite{Wang:2021xih}.

{\it Note added: After this work is finished,~\cite{Geng:2022slq} appears in arXiv, where the authors consider wedge holography with two finite tension branes.}
\begin{acknowledgments}
We are grateful for the useful discussions with our group members in Fudan University. This work is supported by NSFC grant 11905033. YZ is also supported by NSFC 11947301 through Peng Huanwu Center for Fundamental Theory.
\end{acknowledgments}

\appendix

\section{Schwarzian theory on the JT boundary}\label{shw}
In this appendix we show further details of how to compute the JT boundary term and recognize that this is a Schwarzian theory.

We consider the UV cutoff surface $\Sigma$, the tangent vector and normal vector are 
\begin{equation}
    T^{\nu}=\frac{1}{l \cosh(\frac{\rho}{l})} (t',y',0),
\end{equation}
\begin{equation}
    n_{\mu}=(\frac{y'l \cosh\frac{\rho}{l}}{y\sqrt{t'^{2}-y'^{2}}}, -\frac{t'l \cosh \frac{\rho}{l}}{y\sqrt{t'^{2}-y'^{2}}},0).
\end{equation}
The extrinsic curvature is computed as 
\begin{equation}
\begin{split}
    K&=g^{\mu\nu}\nabla_{\mu}n_{\nu}\\&=-\frac{T^{\nu}}{T^{2}}\nabla_{T}n_{\nu}=-\frac{T^{\nu}}{T^{2}}(\partial_{u}n_{\nu}-\Gamma^{\rho}_{\mu\nu}n_{\rho}T^{\mu})\\&=\frac{1}{l \cosh\frac{\rho}{l}} \frac{t'^{3}+y y't''-t'y'^{2}-t'y y''}{(t'^{2}-y'^{2})^{3/2}}
    \end{split}
\end{equation}
By using $y=\epsilon t'$ and expand $K$ to $O(\epsilon^{2})$, we get
\begin{equation}
    K=\frac{1}{l \cosh{\frac{\rho}{l}}}(1-\epsilon^{2}{\rm{Sch}}(t(u),u))
\end{equation}
Where 
\begin{equation}
   {\rm{Sch}}(t(u),u)=\frac{2t't'''-3t''^{2}}{2t'^{2}}.
\end{equation}
\section{Computation of bulk RT surface contribution}\label{RT}
In this appendix, we compute the entropy contribution of RT surface by using embedding coordinates.

The embedding coordinates are
\begin{equation}
\begin{split}
	X^{0}&=\frac{z}{2}+\frac{1}{2 z}\left(\ell^{2}+x^{2}-t^{2}\right) ,\\
	X^{1}&=\frac{\ell^{2}}{z} t ,\\
	X^{2}&=\frac{\ell^{2}}{z} x ,\\
	X^{3}&=\frac{z}{2}-\frac{1}{2 z}\left(\ell^{2}-x^{2}+t^{2}\right),
\end{split}\end{equation}
where $\left(X^{0}\right)^{2}+\left(X^{1}\right)^{2}-\left(X^{2}\right)^{2}-\left(X^{3}\right)^{2}=\ell^{2}$. Using these the geodesic distance $s$ between two points $\left(t_{1}, z_{1}, x_{1}\right)$ and $\left(t_{2}, z_{2}, x_{2}\right)$ is obtained as
\begin{equation}
s=l \operatorname{arccosh}\left[\frac{-\left(t_{2}-t_{1}\right)^{2}+\left(x_{2}-x_{1}\right)^{2}+z_{1}^{2}+z_{2}^{2}}{2 z_{1} z_{2}}\right].
\end{equation}
 Plugging in two points $\text{A}=(t,a\cos\theta,-a\sin\theta)$ and $\text{B}=(t,\epsilon,L)$ where A is the intersection point of DES and EOW brane and B is the right boundary of the interval, one gets
\begin{equation}\begin{split}
\frac{\operatorname{Area}(\Gamma)}{4 G_{N}}
&=\frac{l}{4G_N}\operatorname{arccosh}\left[\frac{(L+a\sin\theta)^2+a^2\cos^2\theta}{2a
	\cos\theta \epsilon}\right].
\end{split}\end{equation}

Considering $\frac{\tilde{\rho}}{\rho_{0}}\ll 1 $  , we can expand the RT result to the first order
\begin{equation}\begin{split}
    \frac{\text{Area}(\Gamma)}{4G_{N}}&=\frac{l}{4G_{N}}\operatorname{arccosh}\left[\frac{(L+a\sin\theta_0)^2+a^2\cos^2\theta_0}{2a \cos\theta_0 \epsilon}\right]\\
    &+\frac{\rho_0\bar{\phi}_r}{4G_{N}a},
\end{split}\end{equation}
where $\frac{1}{\cos\theta}=\cosh(\frac{\rho_{0}+\tilde{\rho}}{l})$ and the solution (\ref{JT}) is used.
\nocite{1}

\bibliography{GDES}

\end{document}